# The compliance of the upper critical field in magic-angle multilayer graphene with the Pauli limit

Evgeny F. Talantsev[1,2]

[1]M.N. Miheev Institute of Metal Physics, Ural Branch, Russian Academy of Sciences, 18, S. Kovalevskoy St., Ekaterinburg, 620108, Russia

[2]NANOTECH Centre, Ural Federal University, 19 Mira St., Ekaterinburg, 620002, Russia

## Abstract

The Pauli limiting field represents fundamental magnetic field at which superconducting state collapses due to the spin-paramagnetic Cooper pair-breaking effect. Cao *et al* [1] reported that the upper critical field (i.e., the magnetic field at which the superconducting state collapses in experiment) in magic-angle twisted trilayer graphene (MATNG, *N*=3) in 2-3 times exceeds the Pauli limiting field in this material. This observation was interpreted as a violation of the Pauli-limiting field in MAT3G. Similar conclusions were recently reported by the same research group in MATNG (*N*=4,5) superlattices [2]. Here we point out that Cao *et al* [1] calculated the Pauli limiting field by the use of reduce (to the weak-coupling limit) full equation of the electron-phonon mediated superconductivity. Considering, that in the same paper, Cao *et al* [1] reported that MATNGs are strong coupled superconductors, we calculate the Pauli limiting field for strong coupled case and show that the observed upper critical fields in MATNGs comply with the Pauli limit. This implies that there is no violation of the Pauli limiting field in the Moiré multilayer graphene superlattices.

The theory of electron-phonon mediated superconductivity describes the ground-state Pauli limiting field (also called Clogston-Chandrasekhar [3,4] limiting field) by the equation [5,6]:

$$B_P(0) = \frac{\Delta(0)}{\sqrt{g}\times\mu_B} \times \left(1 + \lambda_{e-ph}\right) \quad (1)$$

where $\Delta(0)$ is the ground-state superconducting energy gap amplitude, $\mu_B$ is the Bohr magneton, $\lambda_{e-ph}$ is the electron-phonon coupling constant, $g$ is the Lande factor (which is in the absence of spin-orbit scattering equals to $g = 2$).

Cao *et al* [1], and majority of other research groups (extended reference list can be found elsewhere [2,7-9]), used the reduced form of Eq. 1, which is the weak-coupling approximation of full equation for *s*-wave superconducting state of the Bardeen-Cooper-Schrieffer theory (i.e., $\lambda_{e-ph} = 0$, $g = 2$, and $\frac{2\Delta(0)}{k_B T_c} = 3.53$ (where $k_B$ is the Boltzmann constant, $T_c$ is the superconducting transition temperature)):

$$B_{P,BCS}(0) = \frac{\Delta(0)}{\sqrt{g}\times\mu_B} = \frac{1}{\sqrt{8}} \times \frac{k_B}{\mu_B} \times \left[\frac{2\Delta(0)}{k_B T_c}\right] \times T_c = 1.86 \times T_c. \quad (2)$$

Based on Eq. 2, full equation (Eq. 1) can be rewritten in the similar form:

$$B_P(0) = \frac{\Delta(0)}{\sqrt{g}\times\mu_B} \times \left(1 + \lambda_{e-ph}\right) = 1.86 \times \left(1 + \lambda_{e-ph}\right) \times T_c. \quad (3)$$

Eq. 3 shows that the ground-state Pauli limiting field, $B_P(0)$, is enhanced by multiplicative factor of $\left(1 + \lambda_{e-ph}\right)$ vs its BCS counterpart, $B_{P,BCS}(0)$. Considering that the lowest $\lambda_{e-ph} = 0.43$ for electron-phonon mediated superconductors is attributed to aluminium [5], the correct form of Eq. 2 should have the multiplicative factor of 2.65, instead of 1.86. The upper limit of Eq. 3 can be calculated utilizing the highest reported $\lambda_{e-ph}$ value for electron-phonon mediated superconductors, $\lambda_{e-ph} = 3.0$ measured for $Pb_{0.5}Bi_{0.5}$ alloy [5]. Based on Eqs. 1-3, we can conclude that:

$$1.43 \leq \frac{B_P(0)}{B_{P,BCS}(0)} \leq 4.0 \quad (4)$$

Cao *et al* [1] measured the parallel upper critical field, $B_{c2,||}(T)$ (i.e. the upper critical field when the external magnetic field is applied in parallel direction to the film surface), in MATTGs and deduced the ground-state parallel upper critical field $B_{c2,||}(0)$ by utilizing analytical equation of the Ginzburg-Landau (GL) theory:

$$B_{c2,||}(0) = \frac{B_{c2,||}(T)}{\sqrt{1-\frac{T}{T_c}}} \quad (5)$$

Cao *et al* [1] reported that the $B_{c2,||}(0)$ in MATTG exceeds the $B_{P,BCS}(0)$ (Eq. 2) by factor of 2-3. This was interpreted as a violation of the fundamental Pauli limiting field in MATTG. It should be noted, that similar conclusion was made in practically all papers, where $B_{c2,||}(0)$ was reported in atomically thin superconductors [2,7-9].

Cao *et al* [1] introduced so-called the Pauli violation ratio ($PVR$):

$$PVR \equiv \frac{B_{c2,||}(0)}{B_{p,BCS}(0)}. \quad (6)$$

Considering Eqs. 2,3, we can make a conclusion that:

$$PVR \equiv \frac{B_{c2,||}(0)}{B_{p,BCS}(0)} \leq 1 + \lambda_{e-ph} \quad (7)$$

Thus, reported by Cao *et al* (Fig. 2(e) [1]) the violation of Pauli limiting field in MATTG:

$$2.3 \leq PVR \equiv \frac{B_{c2,||}(0)}{B_{p,BCS}(0)} \leq 3.3 \quad (7)$$

and, more recently, by the same group [2], in MAT4G and MAT5G:

$$1.8 \leq PVR \equiv \frac{B_{c2,||}(0)}{B_{p,BCS}(0)} \leq 2.7, \quad (8)$$

actually means that MATNGs (*N*=3,4,5) exhibit:

$$0.8 \leq \lambda_{e-ph} \leq 2.3. \quad (9)$$

This conclusion (Eq. 9) is in a nice agreement with the report by Cao *et al* [1] and Park *et al* [10] who found that the coupling strength in MATNGs (*N*=3,4,5) is varied from weak to ultra-strong level [1,2,10]. Despite Cao *et al* [1] and Park *et al* [2,10] did not provide particular quantitative values that characterize the coupling strength in MATTGs, the range of $\lambda_{e-ph}$ can be estimated by taking into account that elemental niobium (which can be considered as a natural border between weak- and strong-coupled superconductors [5]) exhibits $\lambda_{e-ph} = 1.02 - 1.26$ [5]. Thus, expected range of the $PVR$ is:

$$2.0 \leq PVR \leq 4.0, \quad (10)$$

which is in a nice agreement with reported values by Cao *et al* [1] and Park *et al* [2].

In Figures 1 and 2 we show all $B_{c2,||}(T)$ datasets for MATTGs reported by Cao *et al* [1] where we calculate respectful $\lambda_{e-ph}$ value based on equation:

$$B_{c2,||}(0) = B_P(0) = 1.86 \times \left(1 + \lambda_{e-ph}\right) \times T_c \qquad (11)$$

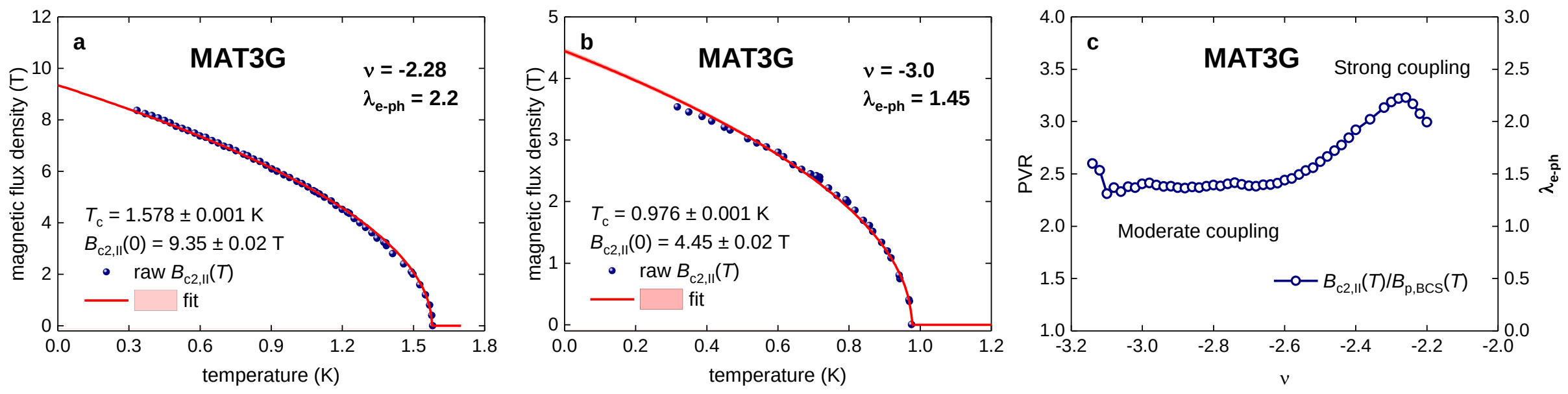

**Figure 1.** The upper critical field, $B_{c2,||}\left(T, \nu, \frac{D}{\varepsilon_0}\right)$, and the data fit to Ginzburg–Landau expression (Eq. 5) (a,b) and PVR (c) in MATTG. All measurements are taken at displacement field $\frac{D}{\varepsilon_0} = -0.41\ \frac{V}{nm}$. The data points for panels **a** and **b** denote constant-resistance at 10% of the zero-field normal-state resistance. **a**, $B_{c2,||}\left(T, \nu, \frac{D}{\varepsilon_0}\right)$ at ν = −2.28 and deduced $\lambda_{e-ph} = 2.2$ (raw data is from Fig. 2,b [1]). **b**, $B_{c2,||}\left(T, \nu, \frac{D}{\varepsilon_0}\right)$ at ν = −3.0 and deduced $\lambda_{e-ph} = 1.45$ (raw data is from Fig. 2,d [1]). **c**, PVR as a function of ν and deduced $\lambda_{e-ph}$ (raw data is from Fig. 2,e [1]).

In the Supplementary Information we show the analysis for $B_{c2,||}(T)$ data in MAT4G reported by Park *et al* [2].

Summarizing all above, we can conclude that there is no Pauli limiting field violation in MATNGs (*N*=3,4,5), and the observed in experiment $B_{c2,||}(T)$ values reflect a fact that these materials are moderate- or strong-coupled superconductors. Primary reason why several research groups made a conclusion that there is the violation of the Pauli limit filed in atomically thin superconductors is that these groups of researchers calculated the Pauli field by the equation, which was derived in the assumption of the weak-coupling interaction in the superconductor.

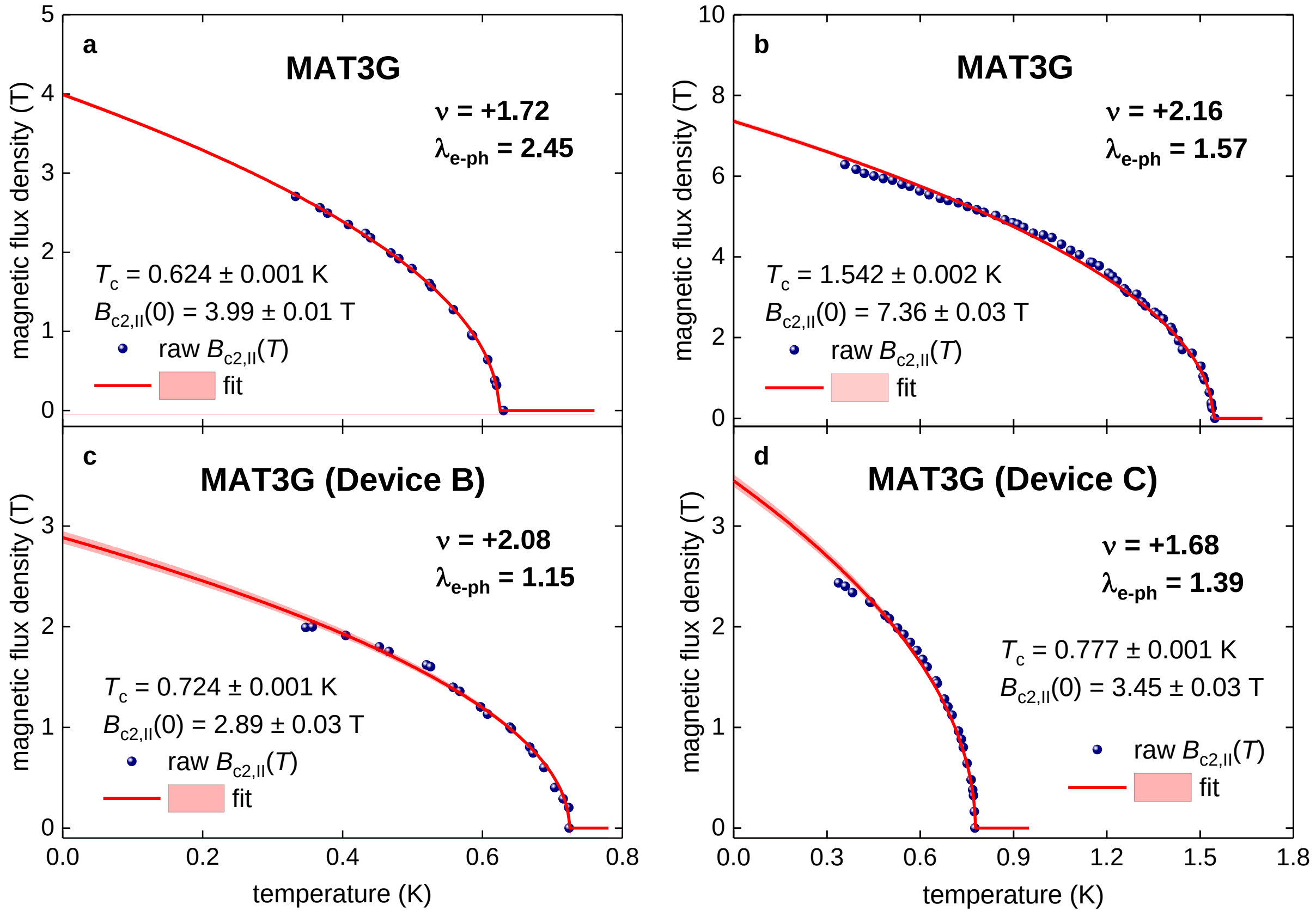

**Figure 2.** The upper critical field, $B_{c2,||}\left(T, \nu, \frac{D}{\varepsilon_0}\right)$, and the data fit to Ginzburg–Landau expression (Eq. 5) in MATTG. The data points denote constant-resistance at 10% of the zero-field normal-state resistance. **a**, $B_{c2,||}\left(T, \nu = 1.72, \frac{D}{\varepsilon_0} = -0.84\ \frac{V}{nm}\right)$ and deduced $\lambda_{e-ph} = 2.45$ (raw data is from Extended Data Fig. 1,a [1]). **b**, $B_{c2,||}\left(T, \nu = 2.16, \frac{D}{\varepsilon_0} = -0.74\ \frac{V}{nm}\right)$ and deduced $\lambda_{e-ph} = 1.57$ (raw data is from Extended Data Fig. 1,b [1]). **c**, $B_{c2,||}\left(T, \nu = 2.08, \frac{D}{\varepsilon_0} = 0.20\ \frac{V}{nm}\right)$ and deduced $\lambda_{e-ph} = 1.15$ (raw data is from Extended Data Fig. 2,a [1]). **d**, $B_{c2,||}\left(T, \nu = 1.68, \frac{D}{\varepsilon_0} = -0.12\ \frac{V}{nm}\right)$ and deduced $\lambda_{e-ph} = 1.39$ (raw data is from Extended Data Fig. 2,b [1]).

**Acknowledgements** The author thanks Jeong Min Park and co-workers of [2] for making raw experimental data is freely available. The author acknowledges financial support provided by the Ministry of Science and Higher Education of Russia (theme “Pressure” No. AAAA-A18-118020190104-3).

**Author contribution** E.F.T. solely contributed to all aspects of this work.

**Competing interests** The author declares no competing interests.

**Data availability statement** No new data was reported or analyzed in the manuscript.

## SUPPLEMENTARY INFORMATION

# The compliance of the upper critical field in magic-angle multilayer graphene with the Pauli limit

Evgeny F. Talantsev[1,2]
[1]M.N. Miheev Institute of Metal Physics, Ural Branch, Russian Academy of Sciences, 18, S. Kovalevskoy St., Ekaterinburg, 620108, Russia
[2]NANOTECH Centre, Ural Federal University, 19 Mira St., Ekaterinburg, 620002, Russia

To represent our statement in more convenient way for the readers, in Figure S1 we show $B_P(0)$ values (calculated by Eq. 1) for MAT4G Device 4C for which the phase diagram is shown by Park *et al* [S1] in their Extended Data Figure 9(c,d). To deduce $B_{c2,||}$ and $T_c$ values from experimental datasets, we utilized the resistance criterion of $R_c = 720 \pm 40\ \Omega$ (which is 10% of the maximum resistance measured for this Device 4C).

It can be seen in Figure S1 that the relation of $B_{c2,||}\left(T{\sim}0.2\ K, \nu, \frac{D}{\varepsilon_0}\right) \cong B_p\left(0, \nu, \frac{D}{\varepsilon_0}, \lambda_{e-ph} = 1.0\right)$ is accurately satisfied across the entire *D*-range. This result implies that the Pauli limit does indeed satisfy in this MAT4G film and this limit perhaps determines the value of the upper critical field in this device.

It should be stressed, that the $B_p\left(0, \nu, \frac{D}{\varepsilon_0}, \lambda_{e-ph}\right)$ values (which are overlapped in Figure S1 with $B_{c2,||}\left(T{\sim}0.2\ K, \nu, \frac{D}{\varepsilon_0}\right)$) were calculated in the assumption of very moderate electron-phonon interaction, $\lambda_{e-ph} = 1.0$. The latter value is slightly lower than the one of pure elemental niobium ($\lambda_{e-ph} = 1.02 - 1.26$ [S2,S3]).

Similar findings (showed in Figure S2) were obtained for the device MAT4G Device 4B (for which the phase diagram is shown by Park *et al* [S1] in their Extended Data Figure 9(a,b)). To deduce $B_{c2,||}$ and $T_c$ values from experimental data, we utilized the resistance criterion of $R_c = 120 \pm 10\ \Omega$ (which is 10% of the maximum resistance measured for this Device 4B).

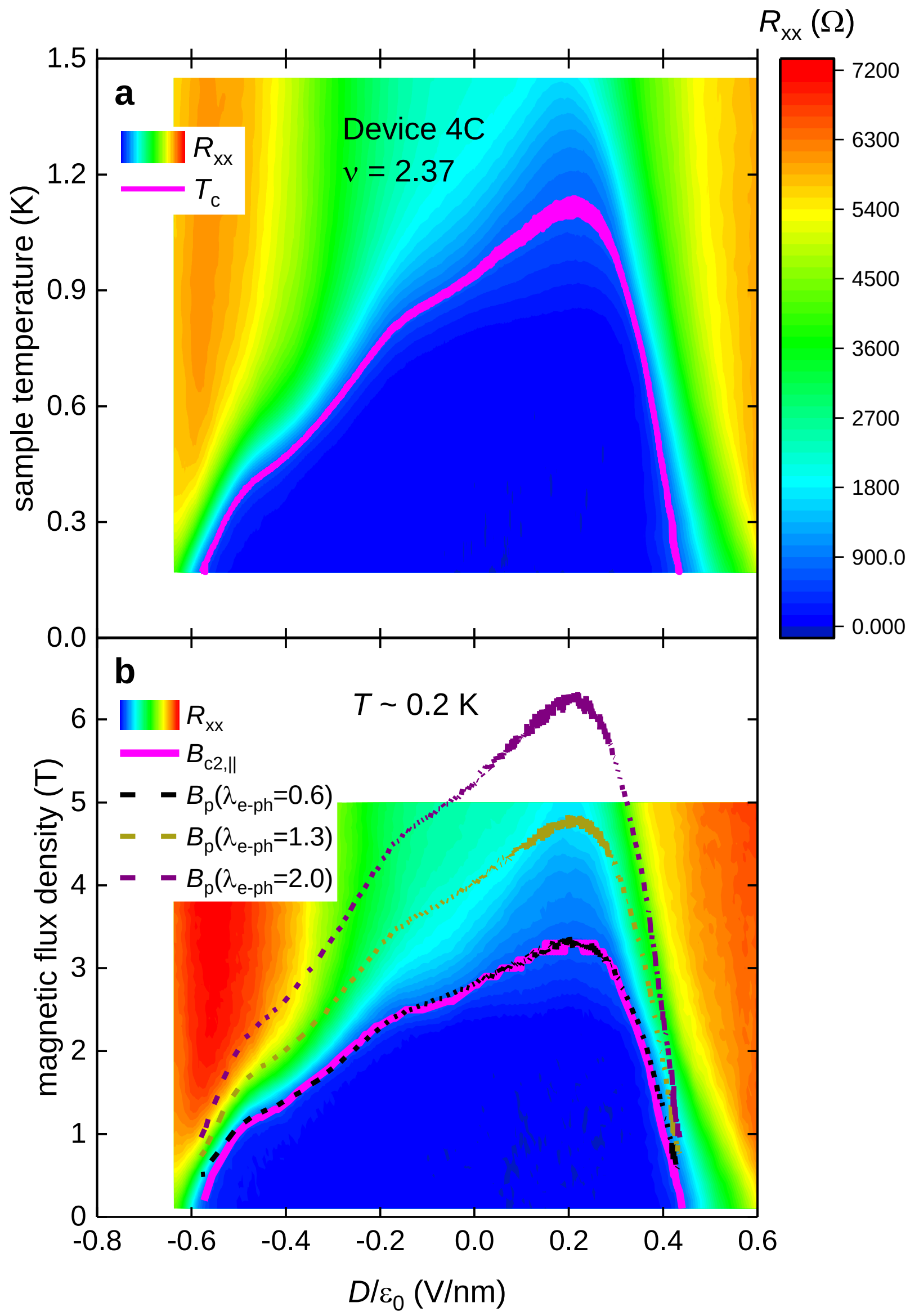

**Figure S1.** $D$-independent compliance of the Pauli limit field, $B_p(0)$ (Eq. 3), with the observed upper critical field, $B_{c2,||}(T\sim0.2\ K)$, in MAT4G. Raw data is from Extended Data Figure 9(c,d) in Ref. S1. $B_{c2,||}$ and $T_c$ values were deduced from experimental data by utilizing the resistance criterion of $R_c = 720\ \pm 40\ \Omega$. (a) $R\left(T, \nu, \frac{D}{\varepsilon_0}\right)$ and deduced $T_c\left(\nu, \frac{D}{\varepsilon_0}\right)$; (b) $R\left(T\sim0.2\ K, B_{||}, \nu, \frac{D}{\varepsilon_0}\right)$, deduced $B_{c2,||}\left(T\sim0.2\ K, \nu, \frac{D}{\varepsilon_0}\right)$, and calculated $B_p\left(0, \nu, \frac{D}{\varepsilon_0}, \lambda_{e-ph}\right)$ (Eq. 3).

It is important to note, that for this device Park *et al* [S1] reported raw $R\left(T\sim0.2\ K, B_{||}, \nu, \frac{D}{\varepsilon_0}\right)$ data (Extended Data Figure 9(a)) and raw $R\left(T, \nu, \frac{D}{\varepsilon_0}\right)$ data (Extended Data Figure 9(b) [S1]) which were measured at slightly different filling factor $\nu = -2.56$ and $\nu = -2.61$, respectively. This implies that $T_c$ values deduced from the Extended Data Figure 9(b) (our Fig. S2(a)) are slightly lower than their counterparts expected for the $\nu = -2.56$

filling factor. However, even for this (favorite for the Pauli limiting field violation) choice of raw data, the inequality of $B_{c2,||}\left(T\sim0.2\ K, \nu, \frac{D}{\varepsilon_0}\right) \leq B_p\left(0, \nu, \frac{D}{\varepsilon_0}, \lambda_{e-ph} = 1.31\right)$ is satisfied across the entire phase diagram (Figure S2).

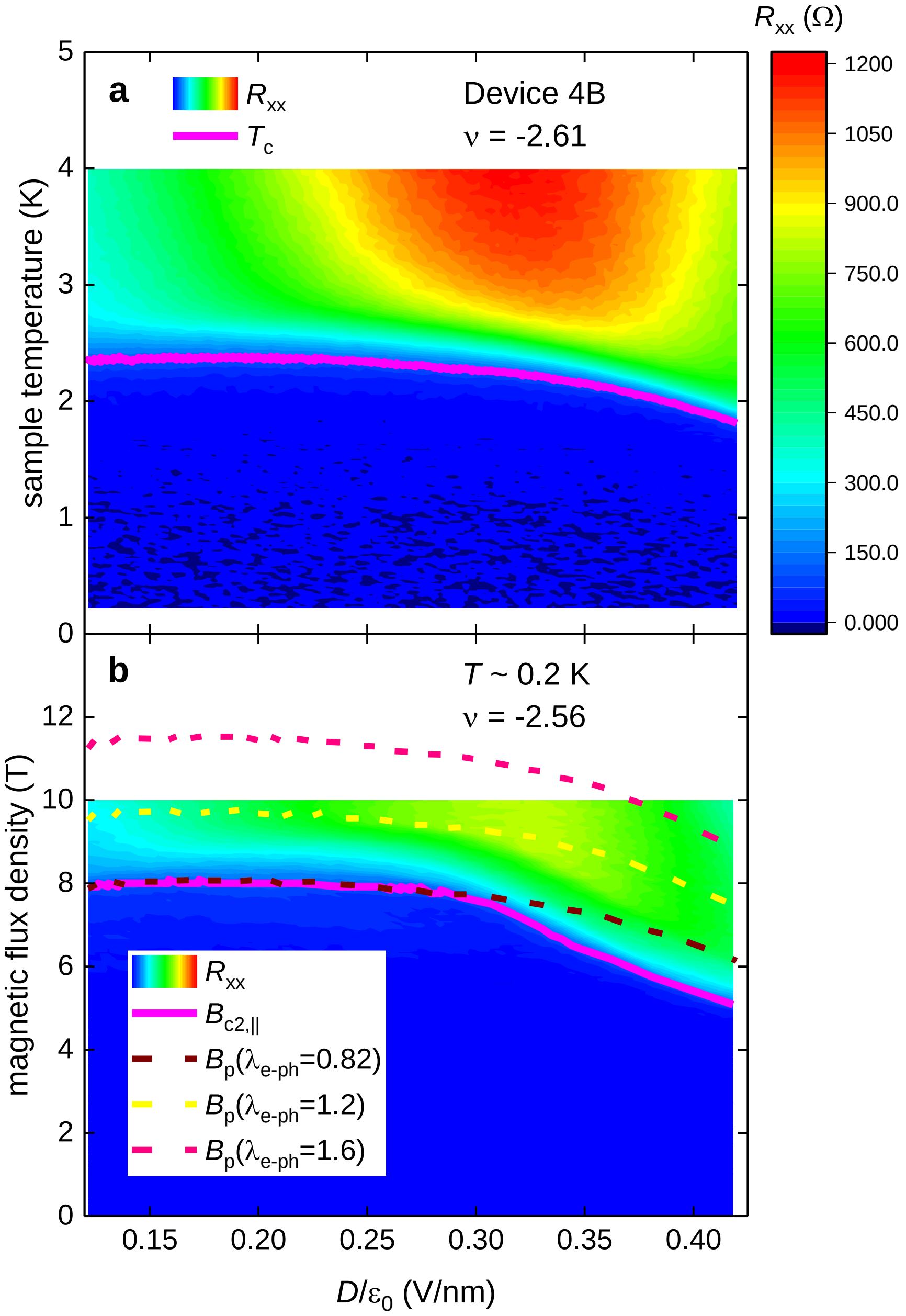

**Figure 2.** *D*-independent compliance of the Pauli limit field, $B_p$ (Eq. 3), with the observed upper critical field, $B_{c2,||}(T\sim0.2\ K)$, in MAT4G (Device 4B). Raw data is from Extended Data Figure 9(a,b) in Ref. S1. $B_{c2,||}$ and $T_c$ values were deduced from experimental data by utilizing the resistance criterion of $R_c = 120\ \pm 10\ \Omega$. (a) $R\left(T, \nu, \frac{D}{\varepsilon_0}\right)$ and deduced $T_c\left(\nu, \frac{D}{\varepsilon_0}\right)$; (b) $R\left(T\sim0.2\ K, B_{||}, \nu, \frac{D}{\varepsilon_0}\right)$, deduced $B_{c2,||}\left(T\sim0.2\ K, \nu, \frac{D}{\varepsilon_0}\right)$, and calculated $B_p\left(0, \nu, \frac{D}{\varepsilon_0}, \lambda_{e-ph}\right)$ (Eq. 3).

It should be stressed that above we calculated $B_p$ (Eqs. S1) and $PVR$ (Eqs. 6,7) in the assumption that MATMGs exhibit *s*-wave symmetry for the superconducting gap. However, this assumption has not been reaffirmed/disproved in any experiment, and if the MATMGs are *d*- or *p*-wave superconductors, then further increase in calculated $B_p(0)$ (Eq. 1) and $PVR$ (Eqs. 6,7) is expected. This is because the weak-coupling limit for *d*-wave case is $\frac{2\Delta(0)}{k_B T_c} = 4.28$ [4], and for *p*-wave case $\frac{2\Delta(0)}{k_B T_c} = 4.06 - 4.92$ [5] (vs $\frac{2\Delta(0)}{k_B T_c} = 3.53$ for *s*-wave [2]). If even MATMGs exhibit *s*-wave gap symmetry, there is a well-established experimental fact that $\frac{2\Delta(0)}{k_B T_c}$ in some unconventional *s*-wave superconductors can be as high as $\frac{2\Delta(0)}{k_B T_c} \geq 9.0$ [6]. In addition, all calculations performed herein were made in the assumption that the Lande factor $g = 2$. However, the values of $g < 2$ are permitted if material exhibits reasonable level of spin-orbit scattering. This implies that calculated $B_p(0)$ (Eq. 1) will be further increased.

The approach presented herein can be equally applied to any superconducting material, especially for atomically thin superconductors [7-9], and for materials where strong-coupling superconductivity is emerged.